\documentclass[prl, showpacs,twocolumn]{revtex4} \usepackage{latexsym}
\usepackage{amsfonts} \usepackage{amsmath,revsymb}

\newcommand{\vp}{{\bf p}}

\newcommand{\vP}{{\bf P}}

\newcommand{\vq}{{\bf q}}

\newcommand{\op}[1]{\hat{#1}}
\newcommand{\vop}[1]{\hat{\bf #1}}

\newcommand{\HS}{{\mathcal{H}}}

\begin{document}

\title{Galilean and Dynamical Invariance of Entanglement in Particle Scattering}

\author{N. L. Harshman}
\affiliation{Department of Computer Science, Audio Technology and Physics\\
4400 Massachusetts Ave., NW, American University, Washington, DC 20016-8058}

\author{S.~Wickramasekara}
\affiliation{Department of Physics, Grinnell College, Grinnell, IA 50112}

\begin{abstract}
Particle systems admit a variety of tensor product structures (TPSs) depending on the algebra of observables chosen for analysis.  
Global symmetry transformations and dynamical transformations may be resolved into local unitary operators with respect to 
certain TPSs and not with respect to others.  Symmetry-invariant and dynamical-invariant 
TPSs are defined and various notions of entanglement are considered for scattering states.
\end{abstract}
\pacs{03.67.Mn, 03.65.Nk, 11.30.-j}
\maketitle

The interaction of particle systems via scattering is a fundamental theoretical and experimental paradigm.  The quantum information theory of particle scattering is, however, still in its infancy.  Results, theoretical and computational, exist for the entanglement between the momenta~\cite{momenta} or the angular momenta~\cite{spin} of two particles generated in scattering, but many problems remain open.  The challenges are partly technical due to the greater complexity of dealing with entanglement in continuous variable systems~\cite{cv} and partly conceptual as in defining a measure of entanglement that has meaningful properties under space-time symmetry transformations.  See, for example, the literature on spin-entanglement of relativistic particles~\cite{relent, goodrel} where different types of entanglement (between two particles, between two particles' spins, and between a single particle's spin and momentum) have been discussed and occasionally confused.

In this letter, we examine how some of these difficulties may be resolved by combining two approaches: (1) the generalized tensor product structures (TPSs) and observable-dependent entanglement developed by Zanardi and others~\cite{zanardiet}, and (2) the representation theory of space-time symmetry groups, which has a long and fruitful history in quantum mechanics.  Using these methods, TPSs for single particle and multi-particle systems are explored.  These methods allow one to distinguish between TPSs that are symmetry invariant and/or dynamically invariant and TPSs that are not, and, in the latter case, to obtain quantitative expressions for the change of entanglement. 
The reason why certain TPSs have entanglement measures which are symmetry or dynamically invariant is that the space-time symmetries or the time evolution operator, respectively, act as a product of local unitaries with respect to these TPSs.

As an application of these general concepts and methods, we will study non-relativistic elastic scattering of two particles.  In this context, several interesting results emerge. First, there are single particle TPSs that are invariant under transformations between inertial reference frames, and these TPSs allow one to define intraparticle entanglement between momentum and spin degrees of freedom in a Galilean invariant manner.  Second, there are multiple, inequivalent two particle TPSs that are symmetry invariant.  In particular, these TPSs can be used to define Galilean invariant entanglement between the internal and external degrees of freedom of the two particle system.  Finally, this internal-external entanglement is also dynamically invariant, i.e., it is conserved during any non-relativistic elastic scattering processes.  

\emph{Single particle TPSs:}  The symmetry group of non-relativistic space-time the Galilean group $\mathcal{G}$ in $3+1$ dimensions. In quantum physics, what is relevant is (the covering group of) $\mathcal{G}$  extended by the central charge of mass.  Let us denote the elements of $\mathcal{G}$ by $g=(b,{\bf a},{\bf v},R)$ where $b\in\mathbb{R}$ is a time translation, ${\bf a}\in\mathbb{R}^3$ a space translation, ${\bf v}\in\mathbb{R}^3$ a velocity boost, and $R\in\mathrm{SO}(3)$ (or $u\in\mathrm{SU}(2)\rightarrow R(u)\in\mathrm{SO}(3)$, the standard 2-to-1 homomorphism) is a rotation.
The associated generators for the unitary representations of $\mathcal{G}$ will be denoted by $\{\op{H},\vop{P},\vop{Q}, \vop{J}\}$.  They form a basis for the Galilei algebra. The mass $\op{M}$ can be added as a central element of this operator Lie algebra. The mass-extended enveloping algebra includes the position operator  $\vop{X}=\vop{Q}\op{M}^{-1}$, the orbital angular momentum vector operator $\vop{L}=\vop{X}\times\vop{P}$, and the intrinsic spin vector operator $\vop{S}=\vop{J}-\vop{L}$.  The operators corresponding to internal energy $\op{W}=\hat{H} - (2\hat{M})^{-1}\vop{P}^2$, intrinsic spin squared $\vop{S}^2$ and (trivially) mass $\hat{M}$ commute with the entire enveloping algebra. They are proportional to the identity operator in the unitary irreducible representations (UIR) of $\mathcal{G}$, $\hat{M}=mI,\ \op{W}=WI,\ \vop{S}^2=s(s+1)I$, and these eigenvalues characterize the representation Hilbert spaces $\HS(m,W,s)$ ~\cite{gal, levy}.
In the single particle case, it is possible to set  $W=0$ without loss of generality.

The UIR Hilbert space $\HS(m,W,s)$ can be always realized as $L^2$-functions defined on the Cartesian product of the spectra of a 
complete system of commuting observables (CSCOs). A variety of such CSCOs exist and choosing one is equivalent to constructing the UIR. 
A standard choice for a CSCO consists of the momentum  vector $\vop{P}$ and the spin component $\op{S}_3$, along with the invariants 
$\hat{M}$, $\op{W}$ and $\vop{S}^2$. 
The Hilbert space $\HS(m,W,s)$ is then realized as
\begin{equation}\label{tps-sp1}
\HS(m,W,s) = \HS_\vp \otimes \HS_s \rightarrow \mathrm{L}^2(\mathbb{R}^3)\otimes \mathbb{C}^{2s+1}.
\end{equation}
The choice of CSCO $\{\vop{P},\op{S}_i\}$ induces the TPS (\ref{tps-sp1}) on the single particle state space, and based on this TPS one can define intraparticle entanglement between momentum degrees of freedom and spin degrees of freedom.

In a general UIR, the action of $U(g)$, $g=(b,{\bf{a}},{\bf{v}},R)\in\mathcal{G}$, on a vector $\phi_\chi(\vp)$ in $\HS(m,W,s)$  is~\cite{levy}
\begin{eqnarray}\label{uir1}
(U(g)\phi)_\chi(\vp) &=& e^{-i\frac{1}{2}m \mathbf{a}\cdot\mathbf{v} +i \mathbf{a}\cdot\vp' -i bE'} \nonumber\\
&&\times \sum_{\chi'}D^{s}({\mathcal{R}}(({\bf p}, E),\tilde{g}))_{\chi'\chi}\phi_{\chi'}(\vp')
\end{eqnarray}
where $\vp' = R\vp + m{\bf v}$, $E = 1/(2m)\vp^2 + W$, $E' = E + {\bf v}\cdot\vp + 1/(2 m){\bf v}^2$, $\tilde{g}=(0,0,{\bf{v}},R)$ 
and $\{{\mathcal{R}}(({\bf p}, E),{\tilde{g}})\}$ is an element of the ``little group" of $\mathcal{G}$ for a massive particle. 
Recall that the little group is the largest subgroup that leaves a standard momentum-energy pair  $({\bf{p}}_0,E_0)$ invariant. 
For a massive particle, the little group of both the Galilean and Poincar\'e groups is isomorphic to the rotation group, and therefore, the 
$D^s({\mathcal{R}}(({\bf{p}},E),\tilde{g})$ is simply the unitary $2s+1$ dimensional representation of the rotation group. 
By definition, the little group, and therewith also the representation \eqref{uir1}, 
depend on the choice of $({\bf{p}}_0,E_0)$, which is arbitrary aside from the constraint $E-\frac{1}{2m}{\bf{p}}^2=W$. 
However, it is known that all of these different representations of $\mathcal{G}$ 
are equivalent~\cite{gal,levy}, and therefore we may use any momentum-energy pair $({\bf{p}},E)$ to construct the general expression for the 
representation. The choice $({\bf{0}},W)$ is particularly simple in that $\{{\mathcal{R}}(({\bf 0}, W),{\tilde{g}})\}=R$, i.e., 
the little group of $\mathcal{G}$ can be chosen to be $SU(2)$, independently of the momentum and energy of the particle. 
With this choice, \eqref{uir1} reduces to the simpler form
\begin{equation}
(U(g)\phi)_\chi(\vp) = e^{-i\frac{1}{2}m \mathbf{a}\cdot\mathbf{v} +i \mathbf{a}\cdot\vp' -i bE'} \sum_{\chi'}D^{s}(R)_{\chi'\chi}\phi_{\chi'}(\vp').
\label{uir1a}
\end{equation} 
The following important property is evident in (\ref{uir1a}): the unitary operators $U(g)$ factor into separate unitary operators $U(g)=U(g)_\vp\otimes U(g)_s$ acting on each Hilbert space in (\ref{tps-sp1}).  In other words, the unitary operators corresponding to Galilean transformations are local unitary operators with respect to the TPS (\ref{tps-sp1}).  Thus, intraparticle entanglement between the spin and the momentum of a free, non-relativistic particle is invariant across inertial reference frames.  We note that this is very different from the relativistic case, where the UIR does not factor and so momentum-spin entanglement is not invariant under coordinate transformations even for free particles~\cite{goodrel}. 

We call a TPS symmetry invariant if there exists a representation $U$ of some group $G$ that factors with respect to this TPS, $U(G)=U_1(G)\otimes U_2(G)\otimes\cdots$.  It is noteworthy that we do not require that each $U_i(G)$ be a nontrivial UIR of the entire group $G$; 
the representation that \eqref{uir1a} furnishes in $\mathbb{C}^{2s+1}$ of \eqref{tps-sp1} has the non-compact part of $\mathcal{G}$ trivially represented by the identity. 

From such a symmetry invariant CSCO one can generally construct new invariant TPSs. For instance, applying a unitary transformation of the form $U=U_1\otimes \mathbb{I}_2$ to the Hilbert space \eqref{tps-sp1} one can get to a new Galilean invariant TPS 
corresponding to the transformed CSCO, $U\{\vop{P},\op{S}_3\}U^{-1}$.
As an example, Fourier transform the first factor to get the CSCO $\{\vop{X},\op{S}_i\}$:
$\mathrm{L}^2(\mathbb{R}^3)\otimes \mathbb{C}^{2s+1}\rightarrow \mathrm{L}^2(\mathbb{R}^3)\otimes \mathbb{C}^{2s+1}.$
Or, transform from rectangular to spherical coordinates to get the CSCO $\{\op{P},\op{\Theta},\op{\Phi}\}$:  
$\mathrm{L}^2(\mathbb{R}^3)\otimes \mathbb{C}^{2s+1}\rightarrow \left(\mathrm{L}^2(\mathbb{R}^+)\otimes \mathrm{L}^2(\mathrm{S}^2)\right)\otimes \mathbb{C}^{2s+1}$.
Finally, choose a CSCO like $\{\op{H},\op{L},\op{L_i}\}$, thereby exploiting harmonic analysis to reduce functions the two-sphere $\mathrm{S}^2$ to an infinite series with the spherical harmonics as basis functions:
$\left(\mathrm{L}^2(\mathbb{R}^+)\right. 
\otimes\left. \mathrm{L}^2(\mathrm{S}^2)\right)\otimes \mathbb{C}^{2s+1} 
\rightarrow \left(\mathrm{L}^2(\mathbb{R}^+)\otimes \left\{\bigoplus_l \mathbb{C}^{2l+1}\right\} \right)\otimes \mathbb{C}^{2s+1}.$
In the TPSs for all of these CSCOs, the part of the UIR that acts on the intrinsic spin Hilbert space can be separated from the part that acts on the rest.

There are, of course, many non-invariant TPS structures of $\HS(m,W,s)$. For example, the TPS induced by the CSCO $\{\op{H},\vop{J}^2, \op{J}_i\}$ can be written as 
$\mathrm{L}^2(\mathbb{R}^+)\otimes \left\{\bigoplus_j \left(\mathbb{C}^{2j+1}\otimes\mathbb{C}^{d(j,s)}\right)\right\}$,
where $d(j,s)$ is a degeneracy parameter that describes the number of times total angular momentum $j$ appears in the coupling of $l$ and $s$.  This TPS is not invariant because the UIR in each subspace depends on variables in the other. 
In general, how a particular TPS transforms under a symmetry group is known when the group representation is well defined with respect to that TPS. Given an entanglement measure defined in terms of the states, it is therefore possible to compute the change of entanglement due to the transformation of states under this symmetry group. In particular, it is possible to compute the time evolution of entanglement for a TPS that does not conserve it by using these techniques.

\emph{Two free particle TPSs:} The total Hilbert space for the two particle states is the tensor product\begin{equation}\label{tps-tp1}
\HS=\HS_A\otimes \HS_B,
\end{equation}
where $\HS_N = \HS(m_N,W_N,s_N),\ N=A,B$.  The representation of $\mathcal{G}$ on $\HS$ factors into a direct product of UIRs.  With respect to the CSCO $\{\vop{P}^A,\op{S}^A_i, \vop{P}^B,\op{S}^B_i\}$, the unitary non-irreducible representation $U({\mathcal{G}})=U_A({\mathcal{G}})\otimes U_B({\mathcal{G}})$ is given by 
\begin{eqnarray}\label{uir2}
&&U(g)\phi(\vp_A,\vp_B)_{\chi_A,\chi_B}\\
&& = e^{-i\frac{1}{2}m_A \mathbf{a}\cdot\mathbf{v} + i\mathbf{a}\cdot\vp_A' - ibE_A'} e^{-i\frac{1}{2}m_B \mathbf{a}\cdot\mathbf{v} + i\mathbf{a}\cdot\vp_B' - ibE_B'}\nonumber\\
&&\times \sum_{\chi_A'\chi_B'}D^{s_A}(R)_{\chi_A'\chi_A}
D^{s_B}(R)_{\chi_B'\chi_B}\phi(\vp_A',\vp_B')_{\chi_A',\chi_B'}\nonumber.
\end{eqnarray}
Note that while the factorization of the representation (\ref{uir2}) with respect to the TPS (\ref{tps-tp1}) implies that 
interparticle entanglement is invariant under Galilean relativity, it does {\em not} 
imply that the vectors $\phi(\vp_A,\vp_B)_{\chi_A,\chi_B}$ are not entangled.  

That the TPS $\HS_A\otimes \HS_B$ is an invariant TPS has nothing to do with the structure of $\mathcal{G}$; rather, it is a general property that holds for any TPS constructed by the direct product of UIR spaces for any group.  In contrast, the invariance of the single non-relativistic particle TPS (\ref{tps-sp1}) is a specific property of the Galilean group.  Using (\ref{tps-sp1}), we find that any partition of the four factors in $\HS = \HS_{\vp_A} \otimes \HS_{s_A}\otimes \HS_{\vp_B}\otimes \HS_{s_B}$
leads to a Galilean invariant TPS.  There are many possibilities, but in particular the bi-partite partition $\HS = \left( \HS_{\vp_A} \otimes \HS_{\vp_B} \right) \otimes \left(  \HS_{s_A} \otimes \HS_{s_B}\right)$
has clear physical relevance: the entanglement between all the momentum degrees of freedom and all the spin degrees of freedom is a Galilean invariant~\cite{bertulani}.

Other symmetry invariant TPSs exist for the two particle case.  For simplicity, consider spinless particles.  
The change of variables to total and relative momentum
\begin{equation}\label{com}
\vP =\vp_A + \vp_B, \quad
\vq = \frac{1}{m_A + m_B} (m_B\vp_A - m_A\vp_B)
\end{equation}
is equivalent to the unitary transformation
\begin{eqnarray}\label{tps-tp4}
\HS_{\vp_A} \otimes \HS_{\vp_B} &\rightarrow& \HS_{\vp} \otimes \HS_{\vq}\nonumber\\
\mathrm{L}^2(\mathbb{R}^3)\otimes\mathrm{L}^2(\mathbb{R}^3) &\rightarrow& \mathrm{L}^2(\mathbb{R}^3)\otimes\mathrm{L}^2(\mathbb{R}^3).
\end{eqnarray}
Applying \eqref{com} to the state $\phi(\vp_A,\vp_B)$ of \eqref{uir2} and using the notation $M=m_A + m_B$ and $\mu = m_Am_B/M$, we find
\begin{equation}\label{uir2b}
U(g)\tilde{\phi}(\vP,\vq) = e^{-i\frac{1}{2}M\mathbf{a}\cdot\mathbf{v} + i\mathbf{a}\cdot\vP' - ibE'} \tilde{\phi}(\vP',\vq'),
\end{equation}
where $\vP' = R\vp + M{\bf v}$, $\vq = R\vq'$, $E =  1/(2M)\vP^2 + W$, $E' = E + {\bf v}\cdot\vP + 1/2 M{\bf v}^2$ and $W= W_A + W_B + 1/(2\mu)\vq^2$.
The only term that depends on $\vq$ is $E'$ and it factors from the rest.  Therefore, (\ref{uir2b}) acts as local unitaries on the TPS $\HS_{\vp} \otimes \HS_{\vq}$ (\ref{tps-tp4}).  This shows that the entanglement between the $\bf{P}$ and $\bf{q}$ degrees of freedom, which we refer to as the internal-external (IE)
entanglement, is Galilean invariant with respect to the TPS \eqref{tps-tp4}.
 When solving the bound state problem, one typically assumes that there is no IE entanglement and so the wave function for the net motion can be factored out from the internal wave function.  As understood in the context of the hydrogen atom~\cite{tommasini98}, having zero internal-external entanglement certainly does not imply that there is no interparticle entanglement, i.e., entanglement with respect to the TPS (\ref{tps-tp1}).  We will discuss this further in the next section.

The transformation of variables (\ref{com}) and of TPS (\ref{tps-tp4}) is the first step in finding the Clebsch-Gordan series for the reduction of the direct product of UIRs of $\mathcal{G}$ to a direct sum~\cite{levy} (partial wave analysis).  One way of writing this direct sum reduction (including spin) is
\begin{eqnarray}\label{galCGC}
&&\HS(m_A,W_A,s_A)\otimes\HS(m_B,W_B,s_B)\\
&&=\int_{W=W_A+W_B}^\infty dW\bigoplus_{j=j_{min}}^\infty \HS(M,W,j)\otimes \mathbb{C}^{d(j,s_A,s_B)}.\nonumber
\end{eqnarray}
There is no sum over mass in the Galilean case, but there is a sum over internal (or center-of-mass) energy $W$ and intrinsic (or total center-of-mass) angular momentum $j$, where $j_{min}=0$ if both particles are either fermions or bosons and $j_{min}=1/2$ otherwise.  Since Galilean group is not simply reducible, the same UIR space $\HS(M,W,j)$ appears a number of times, $d(j,s_A,s_B)$.  
It is the number of ways orbital angular momentum $l$  combines with total spin $s$ to form total angular momentum $j$. The total spin in turn comes from the coupling of $s_A$ and $s_B$.

For particles with spin, the IE TPS (\ref{tps-tp4})  generalizes  as 
\begin{equation}\label{tps-tp5}
\HS(m_A,W_A,s_A)\otimes\HS(m_B,W_B,s_B)=\HS_\vP\otimes \HS_{int}
\end{equation}
where the internal Hilbert space is
\begin{equation}
\HS_{int} = \HS_W \otimes \bigoplus_{j=j_{min}}^\infty \HS(M,W,j)\otimes \mathbb{C}^{d(j,s_A,s_B)}
\end{equation}
The unitary representation of $\mathcal{G}$ on \eqref{tps-tp5} factors $U=U_{\vP}\otimes U_{int}$, where 
\begin{eqnarray}\label{uirext}
U_\vP(g)\psi_e(\vP) = e^{-i(\frac{1}{2}M\mathbf{a}\cdot\mathbf{v} + i\mathbf{a}\cdot\vP' - ibE'_e(\vP)} \psi_e(\vP'),\\
U_{int}(g)\psi_i(W)^{jls}{j_i} = e^{-ibW}\sum_{j_i}D^{j}(R)_{j_i'j_i}\psi_i(W)^{jls}{j'_i},\nonumber
\end{eqnarray}
and $E'_e(\vP) = 1/(2M)\vP^2 + {\bf v}\cdot\vP + 1/2 M{\bf v}^2$. 

\emph{Dynamical invariance of IE entanglement in scattering:}  
We can further extend this result to show that the symmetry invariant TPS \eqref{tps-tp5} is also dynamically invariant.
Partial wave analysis (\ref{galCGC}) allows the use of Shur's lemma which asserts that  an operator in the commutant of a representation is proportional to the identity in every UIR (sub)space. For instance, since the S-operator for elastic scattering is Galilean invariant and unitary, Shur's lemma implies that 
 it acts as the unit operator on each $\HS(M,W,j)$ in (\ref{galCGC}) and as a unitary, symmetric matrix, called the reduced S-matrix,  in each $\mathbb{C}^{d(j,s_A,s_B)}$~\cite{goldberger}.  In the case of a central interaction and spin-orbit coupling, the reduced S-matrix  is just $\exp(2i\delta(W)_{l,s})\delta_{ll'}\delta_{ss'}$, where $\delta(W)_{l,s}$ are called scattering phase shifts.
Clearly, any Galilean invariant S-matrix factors into local unitaries on the IE TPS
(\ref{tps-tp5}).  Therefore, the amount of IE entanglement in an in-state will be invariant under any scattering dynamics that respects Galilean symmetry.

While the amount of IE entanglement for this system under the IE TPS (\ref{tps-tp4}) or (\ref{tps-tp5}) depends on the shape parameters of the input states (see below), it will not be changed by any Galilean invariant dynamics, including all central and non-central, spherically symmetric interactions that depend on relative coordinates. 
The effect of such interactions $\op{V}$ is to change the internal energy of the particle system, $\op{W}=\op{W}_0+\op{V}$, where the subscript $0$ refers to the free particle system. Since internal energy is an invariant in the Galilean algebra, it is possible to define the interacting Hamiltonian $\op{H}=\op{H}_0+\op{V}$ by way of the relation $\op{W}=\op{H}-\frac{1}{2\op{M}}\vop{P}^2$. This amounts to choosing the momentum, angular momentum and boost operators for the interacting system to be the same as those for the free system. If the interaction is spherically symmetric and depends only on the internal variables, then these new interacting generators will fulfill the commutation relations of the Galilei group, and the analysis for the symmetry invariant TPSs can now be carried out verbatim for the interacting case. That is, {\em dynamical invariance is a consequence of symmetry invariance}.  A similar analysis, with some complications, holds for the Poincar\'e group.
Here, the interactions can be included into the invariant mass operator, the relativistic analog of internal energy. However, due to the structure of the Poincar\'e algebra, it is not possible to simply modify the Hamiltonian alone. Different choices of operators that include interactions lead to different forms of dynamics, but the connection between the symmetry invariance and dynamical invariance holds in each case.

\emph{Conclusion:} When considering entanglement in scattering, one generally looks at the interparticle entanglement associated with the TPS \eqref{tps-tp1}.   One assumes that there is no interparticle entanglement in the asymptotic in-state, where the interaction vanishes.  The interparticle entanglement after the  particles emerge from the interaction region can be calculated, for example, using the purity or entropy of the reduced density matrix, where the Hilbert space of one particle has been traced over.  As an example, consider the scattering of spinless particles with Gaussian momentum wavefunctions~\cite{spec2, spec1}:
\begin{equation}\label{eq:g2}
\phi^{in}(\vp_A,\vp_B)= N_Ae^{-\frac{1}{2\sigma_A^2}(\vp_A - \vp_{A0})^2}N_Be^{-\frac{1}{2\sigma_B^2}(\vp_B - \vp_{B0})^2}.
\end{equation}
This in-state has no entanglement under the TPS (\ref{tps-tp1}). However, if we make the variable transformation \eqref{com}, the state $\tilde{\phi}^{in}(\vP,\vq)$ will have entanglement with respect to the IE TPS unless the masses and widths satisfy $\frac{m_A}{\sigma_A^2} =\frac{m_B}{\sigma_B^2}$.
Generally, scattering dynamics transforms the state (\ref{eq:g2}) into an out-state with interparticle entanglement.  However, \cite{spec2} shows that in scattering with a hard-core potential there is no interparticle entanglement in the out-state 
when exactly the same relationship between the masses and widths is satisfied.
Also, it is proved in \cite{spec1} that the wavefunction of a collection of particles with different masses will converge to one with the same mass-width relationship after scattering multiple times, and in this limit, interparticle entanglement tends to zero.  These  results appear to be consequences of the more general principle of dynamical invariance of IE entanglement.

In summary, there exist TPSs that are symmetry invariant, and therefore also dynamically invariant for a large class of potentials. These TPSs allow for measures of entanglement that do not depend on the frame of reference. They further motivate interesting questions about entanglement in particle scattering, such as the 
explicit change of entanglement as a function of time  in the TPS \eqref{tps-tp1} and the dependence of this change on the details of the interaction.

\begin{acknowledgments}

N.L.H.~acknowledges the U.S.-Italian Fullbright Commission and the University of Trento for supporting this work during a sabbatical leave from American University and the Research Corporation for additional support.  S.W.~acknowledges an internal grant from Grinnell College and support from the University of Valladolid where he was a visitor while this work was done.

\end{acknowledgments}

\end{document}